\documentclass[aps,pra,twocolumn,superscriptaddress,showpacs]{revtex4-1}

\usepackage{amsmath}
\usepackage{amsfonts}
\usepackage{amssymb}
\usepackage{graphicx}
\usepackage{color}
\usepackage{subfigure}

\usepackage{hyperref}

\newcommand{\ket}[1]{\left|#1\right\rangle}
\newcommand{\bra}[1]{\left\langle #1\right|}
\newcommand{\abssq}[1]{\left|#1\right|^2}
\newcommand{\expect}[1]{\left\langle #1 \right\rangle}

\begin{document}


\title{Fractional revivals, multiple-Schr\"{o}dinger-cat states and quantum carpets in the interaction of a qubit with $N$ qubits}


\author{Shane Dooley}
\email[]{dooleysh@gmail.com}
\affiliation{Quantum Information Science, School of Physics and Astronomy, University of Leeds, Leeds LS2 9JT, United Kingdom.}


\author{Timothy P. Spiller}
\altaffiliation{York Centre for Quantum Technologies, Department of Physics, University of York, York YO10 5DD, U.K.}
\affiliation{Quantum Information Science, School of Physics and Astronomy, University of Leeds, Leeds LS2 9JT, United Kingdom.}


\date{\today}

\begin{abstract}

We study the dynamics of a system comprised of a single qubit interacting equally with $N$ qubits (a ``spin star'' system). Although this model can be solved exactly, the exact solution does not give much intuition for the dynamics of the model. Here, we find an approximation that gives some insight into the dynamics for a particular class of initial spin-coherent states of the $N$ qubits. We find an effective Hamiltonian for the system that is a finite Kerr (one-axis twisting) Hamiltonian for the $N+1$ qubits. The initial spin coherent state evolves to spin-squeezed states on short time scales, and to ``multiple-Schr\"{o}dinger-cat'' states (superpositions of many spin-coherent states) on longer time scales, a manifestation of the phenomenon of fractional revivals of the initial state. The evolution of the system is visualized with phase-space plots ($Q$ functions) that, when plotted against time, reveal a ``quantum carpet'' pattern. Of particular interest is the fact that our approximation captures the qualitative features of the model even for small values of $N$. This suggests the possibility of observing the phenomenon of fractional revival in this model for systems of few qubits. 


\end{abstract}



\maketitle

\section{Introduction}



Collapse and revival is a well known feature of quantum systems whereby the initial wave packet of the system collapses as it evolves, but at a later time returns (either exactly or approximately) to the initial state: the revival. An example is the evolution of a field mode in a Kerr medium with Hamiltonian $\hat{H}_{k} = \chi_k (\hat{a}^{\dagger}\hat{a})^2$ and initially in a coherent state. After the revival time $T_k = 2\pi/\chi_k$, the field mode is again in the initial state \cite{Yur-86}. \emph{Fractional revival} is an additional effect where at a rational fraction of the revival time the state of the system is made up of a number of superposed, displaced copies of the initial coherent state. For an initial coherent state evolving in a Kerr medium, for example, the field mode will be in a superposition of $q$ coherent states (a ``multiple-Schr\"{o}dinger-cat state'') at a time $T_k/(2q)$ where $q$ is an even number \cite{Gan-91, Tar-93}. 

Another example of collapse and revival is in the Jaynes-Cummings model for the resonant interaction of a field mode and a two-level atom \cite{Ger-05}. The phrase ``collapse and revival'' in this context refers to the collapse and revival of the atomic inversion, but for any initial atom state, the field (initially in a coherent state) returns (approximately) to that coherent state after the revival time. This model also seems to have a limited form of fractional revival: For a judiciously chosen initial atom state, at a quarter of the revival time the field mode is (approximately) in a superposition of two coherent states \cite{Gea-91, Buz-92}. However, this Schr\"{o}dinger-cat state generation is really due to the conditional evolution of the field rather than fractional revival. This is because there are two orthogonal initial atom states that result in two different effective Hamiltonians, $\hat{H}_{JC}^{\pm}= \pm\lambda \sqrt{\hat{a}\hat{a}^{\dagger}}$, for the field evolution \cite{Gea-91}. Starting in a superposition of these two atom states leads to Schr\"{o}dinger-cat states of the field. At no time is the field composed of more than two distinct macroscopic states. 


To see fractional revival and multiple-Schr\"{o}dinger-cat states (with more than two components) in the Jaynes-Cummings model requires sub-Poissonian number statistics for the initial state \cite{Ave-92,Gor-93} (i.e., a non-classical, number-squeezed initial state). In this case one finds the usual collapse and revival, but also ``super-revivals'' at longer times, and multiple Schr\"{o}dinger-cat states at rational fractions of the ``super-revival'' time \cite{Gor-93}. These kinds of fractional revival are again different to the Kerr-type fractional revivals since they come \emph{after} the first revival. Kerr-type fractional revivals, on the other hand, appear \emph{before} the first revival.

The phenomenon of fractional revival is well known and well investigated, both theoretically  \cite{Ave-89, Rob-04} and experimentally \cite{Gre-02}, for systems with infinite-dimensional Hilbert space, including a recent experimental demonstration with a Kerr non-linearity \cite{Kir-13}. There has, however, been less discussion of fractional revival in finite dimensional systems (e.g., a system of $N$ qubits), although Agarwal and co-workers \cite{Aga-97} and Chumakov and co-workers \cite{Chu-99} have studied the evolution of an $N$ qubit system in a ``finite Kerr medium'' analogous to the Kerr Hamiltonian above for the field mode. This type of evolution is also the basis of several proposals to generate Schr\"{o}dinger-cat states of various finite-dimensional systems \cite{Ger-98, Fer-08}. 



In this paper we study the interaction of $N\gg 1$ qubits with a single qubit by an analogue of the Jaynes-Cummings Hamiltonian, assuming throughout an initial spin-coherent state (a ``classical'' state) of the $N$ qubits. In \cite{Doo-13a} it was shown that this system exhibits Jaynes-Cummings-like collapse and revival for a particular class of initial spin coherent states of the $N$ qubits (even for moderate values of $N$). Here we show that the same system with a different class of initial spin coherent states also exhibits \emph{fractional revivals}. These fractional revivals are of the Kerr-type, and are comparable to these found by Agarwal and coworkers and Chumakov and coworkers rather than the Jaynes-Cummings type that requires a sub-Poissonian initial state. The transition from one regime of collapse and revival (Jaynes-Cummings like) to the other (with Kerr-like fractional revival) is made by changing the initial spin-coherent state parameter, something that is in principle very straightforward in the state preparation. We suggest that fractional revivals and the associated multiple-Schr\"{o}dinger-cat states could be observed in this model even for few-qubit systems.

Although we discuss the system of qubits without reference to a particular physical system, there are various candidates for the implementation of the model. For example, as we discuss in the conclusion, the qubits may be superconducting qubits, nuclear spins in symmetric molecules, or spins associated with Nitrogen vacancy (NV) centers in diamond.





\section{Fractional Revivals in Systems of Qubits}

We begin by defining the $J$ operators for an $N$ qubit system as $\hat{J}_{\mu} = \frac{1}{2}\sum_{i = 1}^N \hat{\sigma}_{\mu}^{(i)}$ where $\hat{\sigma}_{\mu}^{(i)}$ is a Pauli $\sigma$-operator for the $i$'th qubit and $\mu \in \{x,y,z\}$. We also define $\hat{J}_{\pm} = \hat{J}_x \pm i \hat{J}_y$ and $\hat{J}^2 = \hat{J}_{x}^2 + \hat{J}_{y}^2+ \hat{J}_{z}^2$. \emph{Dicke states} $\ket{j,m}_N$ are simultaneous eigenstates of $\hat{J}_z$ and $\hat{J}^2$ with eigenvalues $m$ and $j(j+1)$, respectively. In what follows we restrict to the $j=\frac{N}{2}$ eigenspace of the $N$-qubit system. This is the symmetric subspace of the $N$-qubit system: states in this subspace are invariant under permutation of qubits. Alternatively, the system restricted to this subspace can be thought of as a spin-$N/2$ particle. It is an $(N+1)$-dimensional subspace for which the Dicke states $\ket{\frac{N}{2}, m}$ ($m\in\left\{-N/2,\ldots,N/2\right\}$) form a basis. \emph{Spin-coherent states} \cite{Are-72} in this subspace are separable states of the $N$ qubits with each qubit in the same pure state: 
\begin{equation} \ket{\zeta}_N = \left[ \frac{\ket{\downarrow} + \zeta \ket{\uparrow}}{\sqrt{1+\abssq{\zeta}}}\right]^{\otimes N} , \label{eq:SCS} \end{equation} where $\zeta = |\zeta|e^{-i\phi}$ is a complex number. In the Dicke-state basis the spin-coherent state can be written as \begin{equation} \ket{\zeta}_N =  \sum_{m=-N/2}^{N/2} C_m (\zeta) \ket{\frac{N}{2},m}_N ,\label{eq:Dicke}\end{equation} where \begin{equation} C_m (\zeta) = \binom{N}{\frac{N}{2}+m}^{1/2}\frac{\zeta^{\frac{N}{2}+m}}{\left( 1 + \abssq{\zeta} \right)^{N/2}} . \end{equation} By making the transformation $\zeta = e^{-i\phi}\tan\frac{\theta}{2}$ we can also express (\ref{eq:SCS}) in terms of the usual polar and azimuthal Bloch sphere angles $\theta$ and $\phi$ as follows:

\begin{equation} \ket{\zeta}_N = \ket{\theta,\phi}_N = \left(  \cos\frac{\theta}{2}\ket{\downarrow} + e^{-i\phi}\sin\frac{\theta}{2}\ket{\uparrow}  \right)^{\otimes N} . \end{equation}

We note that the spin-coherent states are widely regarded as ``classical'' states of an $N$ spin system \cite{Mar-03, Gir-08} and that the preparation of any spin coherent state $\ket{\zeta_0}_N$ from an initial spin coherent state (e.g., the ground state $\ket{\zeta=0}_N = \ket{\downarrow}^{\otimes N}$) is, in principle, straightforward: the arbitrary spin-coherent state can be generated by an appropriate interaction with an external classical field \cite{Are-72}.


\vspace{-3mm}

\subsection{Interaction of $N$ qubits with one qubit}

We say that a system of $N$ qubits interacts with a single qubit via Hamiltonian \begin{equation} \hat{H} = \omega\hat{J}_{z}  +  \frac{\Omega}{2}\hat{\sigma}_z + \lambda\left( \hat{J}_{+} \hat{\sigma}_{-} +  \hat{J}_{-} \hat{\sigma}_{+} \right) . \label{eq:H} \end{equation} This Hamiltonian was studied in \cite{Hut-04,Bre-04,ElO-11, Wu-14, Doo-13a} and has been referred to as a ``spin star'' system.


Given any initial state of this system we can find an exact expression for the state at any later time $t$. This expression is, however, cumbersome and difficult to interpret in its exact form. Here, for convenience, we focus on the resonance ($\Omega=\omega$) and initial states that are of the form

\begin{equation}  \ket{\Psi (0)} =  \ket{\zeta_0}_N \otimes \left(  \alpha \ket{\phi_0^+} + \beta\ket{\phi_0^-}  \right) , \label{eq:initialstate}  \end{equation} where the $N$-qubit system is in a spin-coherent state and the single qubit is initially in an arbitrary pure state, written here in terms of the orthonormal basis states $\ket{\phi_0^{\pm}} = \frac{1}{\sqrt{2}}\left(\ket{\downarrow}\pm e^{-i\phi_0}\ket{\uparrow} \right)$ that depend on the phase $\phi_0$ of the spin-coherent state parameter $\zeta_0 = |\zeta_0|e^{-i\phi_0}$. We consider separately the evolution of the two orthonormal states $\ket{\Psi^{\pm}(0)}=\ket{\zeta_0}_N \ket{\phi_0^{\pm}}$ by our Hamiltonian since the evolution of an arbitrary initial state (\ref{eq:initialstate}) is just a superposition of these two solutions.

On resonance it is convenient to first rotate Hamiltonian (\ref{eq:H}) to an interaction picture where the Hamiltonian is $\hat{H}_\textup{int} =   \lambda \left( \hat{J}_{-} \hat{\sigma}_{+} + \hat{J}_{+}\hat{\sigma}_{-}  \right)$ (the time independence of this interaction Hamiltonian depends on the resonance condition). The unitary time evolution operator is thus $\hat{U}(t)  = e^{-it \lambda \left( \hat{J}_{-} \hat{\sigma}_{+} + \hat{J}_{+}\hat{\sigma}_{-}  \right) }$. Expanding the exponential as a Taylor series, and using the identities $\hat{\sigma}_{+}\hat{\sigma}_{+} = \hat{\sigma}_{-}\hat{\sigma}_{-} = 0$, $\hat{\sigma}_{+}\hat{\sigma}_{-} = \ket{\uparrow}\bra{\uparrow}$ and $\hat{\sigma}_{-}\hat{\sigma}_{+} = \ket{\downarrow}\bra{\downarrow}$, gives \begin{eqnarray}  \hat{U}(t) &=& \cos\left( \lambda t \sqrt{\hat{J}_{-}\hat{J}_{+}}  \right) \ket{\uparrow}\bra{\uparrow} + \cos\left( \lambda t \sqrt{\hat{J}_{+}\hat{J}_{-}}  \right) \ket{\downarrow}\bra{\downarrow}  \nonumber\\ & &  - i   \sin\left( \lambda t \sqrt{\hat{J}_{+}\hat{J}_{-}} \right) \left( \hat{J}_{+}\hat{J}_{-}  \right)^{-1/2}  \hat{J}_{+} \ket{\downarrow}\bra{\uparrow} \nonumber \\ & & - i  \sin\left( \lambda t \sqrt{\hat{J}_{-}\hat{J}_{+}} \right)  \left( \hat{J}_{-}\hat{J}_{+}  \right)^{-1/2} \hat{J}_{-} \ket{\uparrow}\bra{\downarrow} .  \label{eq:Ua} \end{eqnarray} Depending on whether the initial state of the qubit is $ \ket{\phi_0^{+}}$ or $ \ket{\phi_0^{-}}$ we write this unitary operator as $\hat{U}^{+}(t)$ or $\hat{U}^{-}(t)$, i.e., $\hat{U}(t) = \hat{U}^{+}(t) + \hat{U}^{-}(t)$ with $\hat{U}^{\pm}(t) \equiv U(t) \ket{\phi_0^{\pm}}\bra{\phi_0^{\pm}}$. From (\ref{eq:Ua}) we find that

\begin{widetext}
\begin{eqnarray}  \hat{U}^{\pm}(t) &=& \Bigg[ \frac{1}{\sqrt{2}} \,  \cos\left( \lambda t \sqrt{\hat{J}_{+}\hat{J}_{-}}  \right) \ket{\downarrow} \mp i \frac{ e^{-i\phi_0}}{\sqrt{2}} \, \sin\left( \lambda t \sqrt{\hat{J}_{+}\hat{J}_{-}} \right)  \left( \hat{J}_{+}\hat{J}_{-}  \right)^{-1/2}  \hat{J}_{+} \ket{\downarrow} \nonumber\\ && \pm \frac{ e^{-i\phi_0}}{\sqrt{2}} \, \cos\left( \lambda t \sqrt{\hat{J}_{-}\hat{J}_{+}} \right) \ket{\uparrow} - \frac{i}{\sqrt{2}} \, \sin\left( \lambda t \sqrt{\hat{J}_{-}\hat{J}_{+}} \right) \left( \hat{J}_{-}\hat{J}_{+}  \right)^{-1/2} \hat{J}_{-}  \ket{\uparrow} \Bigg] \bra{\phi_0^{\pm}} . \label{eq:U}\end{eqnarray} 
\end{widetext}

This is still an exact expression that does not give much insight into the features of the system as it evolves in time. However, we can use two approximations that allow us to write the time-evolved state in a useful way.

\emph{First approximation}. We say that the spin-coherent state $\ket{\zeta_0}_N$ is approximately an eigenstate of the operator $\left( \hat{J}_{-}\hat{J}_{+}  \right)^{-1/2} \hat{J}_{-}$ with complex eigenvalue $e^{-i\phi_0}$ and an eigenstate of $\left( \hat{J}_{+}\hat{J}_{-}  \right)^{-1/2}  \hat{J}_{+}$ with eigenvalue $e^{i\phi_0}$. 

This is a good approximation when (details given in Appendix \ref{app:a}):

\begin{equation}  \frac{1}{\sqrt{N}} \ll \left| \zeta_0 \right| \ll \sqrt{N} \quad \mbox{and} \quad 1 \ll N . \label{eq:approx1} \end{equation} For large $N$ this restriction is not at all severe since a very broad range of values of $\zeta_0$ will satisfy condition (\ref{eq:approx1}).

\begin{figure}
\includegraphics[width=85mm, scale=0.62]{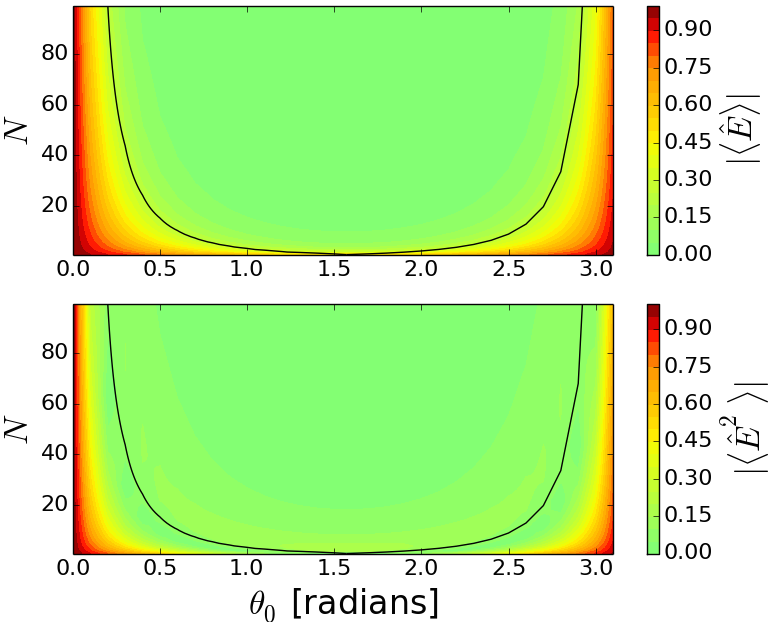}
\caption{(Color online). Top: $\left|\bra{\zeta_0}\hat{E}\ket{\zeta_0}\right|$ plotted against $\theta_0 = 2\arctan |\zeta|$ and $N$ for $\phi_0 = 0$. Bottom: $\left|\bra{\zeta_0}\hat{E}^2\ket{\zeta_0}\right|$ plotted against $\theta_0$ and $N$ for $\phi_0 = 0$. The black lines show $\abssq{\zeta_0}=1/N$ for $|\zeta_0|<1$ (or $\theta_0 < \pi/2$) and $\abssq{\zeta_0}=N$ for $|\zeta_0|>1$ (or $\theta_0 > \pi/2$).}
\label{fig:approx1}
\end{figure}

We plot in Fig. \ref{fig:approx1} the quantities $\left| \bra{\zeta_0}\hat{E}\ket{\zeta_0}\right|$ and $\left| \bra{\zeta_0}\hat{E}^2\ket{\zeta_0}\right|$ where $\hat{E} \equiv e^{-i\phi_0} - \left( \hat{J}_{-}\hat{J}_{+}  \right)^{-1/2} \hat{J}_{-}$. The fact that both of these quantities are small when $ \frac{1}{\sqrt{N}} \ll |\zeta_0|  \ll \sqrt{N}$ justifies our approximation.

We note that a similar approximation can be made for a field mode where $\left( \hat{a}\hat{a}^{\dagger} \right)^{-1/2} \hat{a}\ket{\alpha} \approx e^{-i\phi}\ket{\alpha} $ for $\ket{\alpha}=\ket{|\alpha|e^{-i\phi}}$ a coherent state with $|\alpha| \gg 1$ \cite{Lou-73}.

Using the identities $\hat{J}_{+}\hat{J}_{-} = \hat{J}^2 - \hat{J}_z^2 + \hat{J}_z$ and $\hat{J}_{-}\hat{J}_{+} = \hat{J}^2 - \hat{J}_z^2 - \hat{J}_z$ then allows us to write (\ref{eq:U}) as

\begin{equation}  \hat{U}^{\pm}(t) = e^{\mp i \lambda t \sqrt{ \hat{J}^2 - \hat{J}_z^2 - \hat{J}_z \otimes \hat{\sigma}_z }} \ket{\phi_0^{\pm}}\bra{\phi_0^{\pm}} . \end{equation} In other words, the effective Hamiltonian given the initial qubit state $\ket{\phi_0^{\pm}}$ is  \begin{equation} \hat{H}^{\pm} =  \pm \lambda \sqrt{\hat{J}^2 - \hat{J}_z^2 - \hat{J}_z \otimes \hat{\sigma}_z} . \label{eq:Heff2}\end{equation} Unlike Hamiltonian $\hat{H}$ this effective Hamiltonian is diagonal in the Dicke-state basis of the $N$ qubits, in the sense that $\bra{\frac{N}{2},m} \hat{H}^{\pm}\ket{\frac{N}{2},m'} \propto \delta_{mm'}$.


The fact that we can find an effective Hamiltonian given the initial qubit state $\ket{\phi^{+}}$ or $\ket{\phi^{-}}$ is reminiscent of the effective Hamiltonian that can be derived in the strong field approximation of the resonant Jaynes-Cummings model: when the atom is in the state $\ket{\phi^{\pm}}$, a ``semi-classical eigenstate,'' the field mode evolves by the effective Hamiltonian $\hat{H}_\textup{JC}^{\pm}= \pm\lambda \sqrt{\hat{a}\hat{a}^{\dagger}}$ \cite{Gea-91, Chu-94}. Indeed, in the parameter regime $\frac{1}{\sqrt{N}}\ll |\zeta_0| \ll 1$ the above model exhibits Jaynes-Cummings-like collapse and revival, as discussed in \cite{Doo-13a}. Making the transformations $\lambda \to \lambda/\sqrt{N}$ and $\zeta \to \zeta/\sqrt{N}$ and taking $N\to\infty$, Hamiltonian (\ref{eq:H}) reduces exactly to the Jaynes-Cummings model \cite{Hol-40} and the initial spin-coherent state (\ref{eq:Dicke}) becomes a coherent state \cite{Mar-03,Doo-13a}. 


\emph{Second approximation}. Expanding the square root in (\ref{eq:Heff2}) in powers of the operator $\hat{M} = \frac{ \hat{J}_z^2 + \hat{J}_z \otimes \hat{\sigma}_z }{\hat{J}^2}$ we obtain

\begin{equation} \hat{H}^{\pm} = \pm \lambda \sqrt{\hat{J}^2 } \: \sum_{k=0}^{\infty} A_k \hat{M}^k ,\end{equation} where $A_0 =1$, $A_1 = -1/2$ and $A_k = -(2k-3)!!/(2^k k!)$ for $k\geq 2$ are coefficients whose absolute value is always less than unity. [Here $(2k-3)!!$ is a double factorial, i.e., the product of all odd positive integers less than or equal to $2k-3$.] The second approximation is to truncate this expression to the first two terms:

\begin{equation} \hat{H}_\textup{trunc}^{\pm} = \pm \lambda \left( \sqrt{ \hat{J}^2  } - \frac{ \hat{J}_z^2 +  \hat{J}_z \otimes \hat{\sigma}_z }{ 2 \sqrt{ \hat{J}^2  }} \right) . \label{eq:Hefftrunc}  \end{equation} 

This approximation is valid when $ |\zeta_0| \approx 1$ (or $\theta_0 \approx \pi/2$) and $\lambda t \ll N$ (details given in Appendix \ref{app:b}). In Fig. \ref{fig:precision} we plot $|\langle \tilde{\Psi}(t) | \Psi (t) \rangle|$, the fidelity of the exact state to the approximation against time for various initial values of $\theta_0$ and $N$. The time axis is scaled by $T=\frac{2\pi}{\lambda}\sqrt{\frac{N}{2}\left( \frac{N}{2} + 1 \right)}$ for reasons that should become clear in the next section.

\begin{figure}
\includegraphics[width=85mm, scale=0.62]{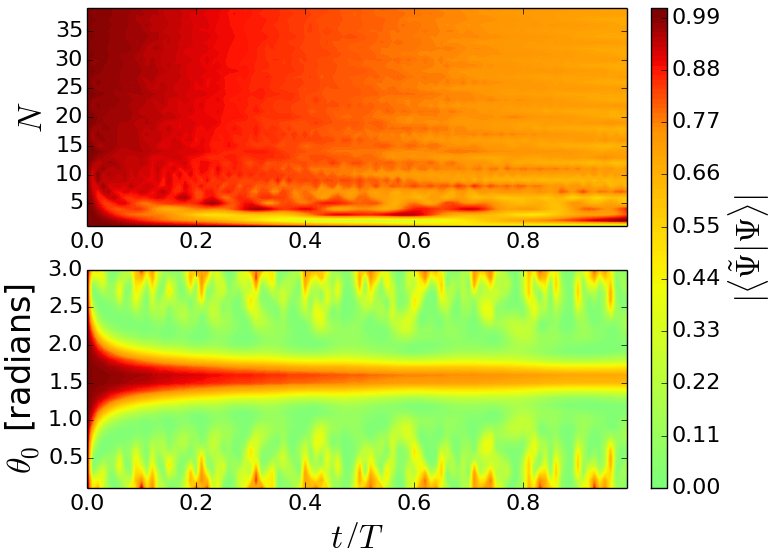}
\caption{(colour online). Top: Fidelity of approximation to the exact state plotted against $N$ and $t/T$ with $\theta_0=\pi/2$ and $\phi_0 = 0$. Bottom: Fidelity plotted against $\theta_0$ and $t/T$ with $N=40$ and $\phi_0 =0$. For $\theta_0 \approx \pi/2$ (or $|\zeta_0| \approx 1$) and $\lambda t \ll N$ (or $t\ll T$) the fidelity is high (red).}
\label{fig:precision}
\end{figure}


The truncated effective Hamiltonian (\ref{eq:Hefftrunc}) can be conveniently rewritten as

\begin{equation}  \hat{H}_\textup{trunc}^{\pm} = \pm \lambda \left( \sqrt{ \hat{J}^2 } + \frac{1}{8 \sqrt{ \hat{J}^2 } } - \frac{ \left( \hat{J}_z +  \frac{\hat{\sigma}_z}{2} \right)^2}{ 2 \sqrt{ \hat{J}^2 }} \right) . \label{eq:HeffOAT} \end{equation}

If the initial state is either $\ket{\Psi^+ (0)}$ or $\ket{\Psi^{-}(0)}$ (rather than some superposition of the two) then (since both $\ket{\Psi^+ (0)}$ and $\ket{\Psi^{-}(0)}$ are eigenstates of $\hat{J}^2$) the first two terms of (\ref{eq:HeffOAT}) just give a global phase factor that can be ignored. If the initial state is a superposition of $\ket{\Psi^+ (0)}$ and $\ket{\Psi^{-}(0)}$ then the first two terms cannot be ignored since they give a relative phase factor. The last term of (\ref{eq:HeffOAT}) is proportional to $\left( \hat{J}_z + \frac{\sigma_z}{2} \right)^2$. For an $N$-qubit system the Hamiltonian $\hat{H}_\textup{oat} = \chi_\textup{oat}\hat{J}_z^2$ is known as a \emph{one-axis twisting} Hamiltonian \cite{Kit-93,Ma-11} or, because it is analogous to the Kerr Hamiltonian for a field mode -- a \emph{finite Kerr} Hamiltonian \cite{Chu-99}. Our term proportional to $\left( \hat{J}_z + \frac{\sigma_z}{2} \right)^2$ in the effective Hamiltonian (\ref{eq:HeffOAT}) is therefore a one-axis twisting term for the $(N+1)$-qubit system. 

Previous examples of finite-dimensional systems whose Hamiltonians include one-axis twisting terms are a collection of two-level atoms interacting with a far detuned field mode \cite{Aga-97,Kli-98}; Bose-Einstein condensates in a double-well potential \cite{Mil-97}; molecular nano-magnets \cite{Wer-08}; a collection of NV centers coupled to the vibrational mode of a diamond resonator \cite{Ben-13}. Here we have shown that the one-axis-twisting Hamiltonian can also be an effective Hamiltonian in the interaction of $N$ qubits with a single qubit. We note, however, that the coupling parameter $\lambda$ is weakened by a factor of $1/(2\sqrt{\hat{J}})$ for the one-axis-twisting term in $\hat{H}_\textup{trunc}^{\pm}$.

\subsection{Multiple Cat States and Quantum Carpets}


In this subsection (following the method of analysis of \cite{Ave-89, Rob-04}) we investigate the evolution of the initial state $\ket{\Psi^{\pm} (0)} = \ket{\frac{\pi}{2},\phi_0}_N \ket{\phi_0^{\pm}}$ under the approximations made in the previous section. We note that the initial spin coherent state has $\theta_0 = \pi/2$ (or, equivalently, $|\zeta_0|=1$) so that all of our approximation conditions are satisfied so long as $\lambda t \ll N$ and $1\ll N$. 






The system evolves by the effective truncated Hamiltonian $\hat{H}_\textup{trunc}^{\pm}$ so that the evolved state is

\begin{equation} \ket{\tilde{\Psi}^{\pm}(t)} = \sum_{m=-\frac{N}{2}}^{N/2}  \frac{C_m}{\sqrt{2}}  \ket{\frac{N}{2},m} \left( F_m^{\pm}(t)\ket{\downarrow} \pm e^{-i\phi_0} G_m^{\pm}(t) \ket{\uparrow} \right) , \label{eq:evolved}  \end{equation} where the tilde above $\Psi$ indicates approximation and where we define 


\small \begin{eqnarray} F_m^{\pm}(t) \hspace{-0.5mm} &=& \hspace{-0.5mm} \exp\hspace{-0.5mm}\left[\mp it \lambda\hspace{-0.5mm} \left(\hspace{-1.0mm} \sqrt{\frac{N}{2}\left( \frac{N}{2} + 1 \right)} - \hspace{-0.5mm}\frac{m(m-1)}{2\sqrt{\frac{N}{2}\left( \frac{N}{2} + 1 \right)}} \right)\hspace{-0.5mm} \right] , \label{eq:F} \\  G_m^{\pm}(t)\hspace{-0.5mm} &=&\hspace{-0.5mm} \exp\hspace{-0.5mm}\left[\mp it \lambda\hspace{-0.5mm} \left( \hspace{-1.0mm}\sqrt{\frac{N}{2}\left( \frac{N}{2} + 1 \right)} - \hspace{-0.5mm}\frac{m(m+1)}{2\sqrt{\frac{N}{2}\left( \frac{N}{2} + 1 \right)}} \right)\hspace{-0.5mm} \right] . \label{eq:G} \end{eqnarray}\normalsize So that we can eventually arrive at a more useful expression for $\ket{\tilde{\Psi} (t)}$ we now consider some properties of these functions $F_m^{\pm}(t)$ and $G_m^{\pm}(t)$. First, when $N$ is an even number both $F_m^{\pm}(t)$ and $G_m^{\pm}(t)$ are periodic in time with period $T=\frac{2\pi}{\lambda}\sqrt{\frac{N}{2}\left( \frac{N}{2} + 1 \right)}$. To see this note that when $N$ is even both $\frac{N}{2}\left( \frac{N}{2} + 1 \right)$ and $m(m\pm1)$ are even integers so that $F_m^{\pm}(T)$ and $G_m^{\pm}(T)$ are exponentials whose phases are integer multiples of $2\pi$. Since $F_m^{\pm}(t)$ and $G_m^{\pm}(t)$ are periodic in time we have $\ket{\tilde{\Psi}^{\pm}(t+T)} = \ket{\tilde{\Psi}^{\pm}(t)}$: the system returns to its initial state after period $T$. Similarly, when $N$ is an odd number we have $\ket{\tilde{\Psi}^{\pm}(t+2T)} = \ket{\tilde{\Psi}^{\pm}(t)}$ and the state has a revival time of $2T$.

Focussing on the case when $N$ is even, we now consider times $t= \frac{pT}{q}$ where $p$ and $q$ are coprime integers, i.e., rational fractions of the revival time. Then $F_m^{\pm}\left(pT/q\right)$ and $G_m^{\pm}(pT/q)$ are both either periodic functions of the discrete variable $m$ with period $q$ if $q$ is odd, or anti-periodic functions of $m$ with (anti-) period $q$ if $q$ is even:

\begin{eqnarray}  F_m^{\pm} &=& F_{m+q}^{\pm}\: ; \quad  G_m^{\pm} = G_{m+q}^{\pm}  \quad (q \: \mbox{odd}), \\  F_m^{\pm} &=& -F_{m+q}^{\pm}\:  ; \quad  G_m^{\pm} = -G_{m+q}^{\pm}  \quad (q \: \mbox{even}) . \end{eqnarray} Either way, this means that we can write $F_m^{\pm}(pT/q)$ and $G_m^{\pm}(pT/q)$ in terms of their discrete Fourier transforms, \begin{equation}   \mathcal{F}_l^{\pm} =  \frac{1}{\sqrt{q}} \sum_{m=0}^{q-1} F_m^{\pm} \, e^{i\phi_l m} \: ; \quad   {G}_l^{\pm} = \frac{1}{\sqrt{q}} \sum_{m=0}^{q-1} G_m^{\pm} \, e^{i\phi_l m} ,  \label{eq:fourier2} \end{equation} where we define \begin{equation}  \phi_l \equiv \begin{cases} 2\pi l / q   &   \text{if } q  \text{ is odd},  \\    \pi (2l+1)/ q   &   \text{if } q  \text{ is even} . \end{cases}  \end{equation} The inverse transform is

\begin{equation}  F_m^{\pm} = \frac{1}{\sqrt{q}} \sum_{l=0}^{q-1} \mathcal{F}_l^{\pm} \, e^{-i\phi_l m} \: ; \quad   G_m^{\pm} = \frac{1}{\sqrt{q}} \sum_{l=0}^{q-1} \mathcal{G}_l^{\pm} \, e^{-i\phi_l m} . \label{eq:fourier1} \end{equation} 

The advantage of writing the functions in terms of their discrete Fourier transforms is that the phases in the exponentials $F_m^{\pm}$ and $G_m^{\pm}$ are now linear in $m$ rather than quadratic in $m$. Substituting (\ref{eq:F}) and (\ref{eq:G}) into (\ref{eq:fourier2}) and using the fact that $e^{i\phi_l m}F_m^{\pm}(pT/q)$ and $e^{i\phi_l m}G_m^{\pm}(pT/q)$ are periodic in $m$, it is not difficult to show that \begin{equation}   \mathcal{G}_l^{\pm} =  e^{-i\phi_l} \mathcal{F}_l^{\pm}. \label{eq:samefourier}\end{equation} Substituting (\ref{eq:fourier1}) and (\ref{eq:samefourier}) into our expression (\ref{eq:evolved}) allows us to write the state $\ket{\Psi^{\pm} (t)}$ appealingly as: \begin{eqnarray}  \ket{\tilde{\Psi}^{\pm} (pT/q)} = \frac{1}{\sqrt{q}} \sum_{l=0}^{q-1} \mathcal{F}_l^{\pm} e^{i\phi_l N/2} \ket{\frac{\pi}{2}, \phi_0 + \phi_l}_N  \nonumber\\  \otimes \frac{1}{\sqrt{2}} \left( \ket{\downarrow} \pm e^{-i(\phi_0 + \phi_l)} \ket{\uparrow} \right) . \label{eq:fracrev} \end{eqnarray} This is a superposition of $q$ terms involving spin-coherent states $\ket{\frac{\pi}{2}, \phi_0 + \phi_l}_N$ of the $N$-qubit system distributed uniformly in the azimuthal Bloch sphere angle $\phi$. Expression (\ref{eq:fracrev}) indicates that the system undergoes fractional revivals at times $t=pT/q$ since it shows that the state at that time is a superposition of displaced copies of the initial wave packet. This is seen most clearly by rewriting (\ref{eq:fracrev}) as $ \ket{\tilde{\Psi}^{\pm}(pT/q)} = \hat{U}^{\pm}_\textup{trunc}(pT/q)\ket{\Psi^{\pm}(0)}$ where \begin{eqnarray}   \hat{U}^{\pm}_\textup{trunc}(pT/q) &=& e^{-ipT\hat{H}_\textup{trunc}^{\pm}/q} \nonumber\\ &=&  \frac{1}{\sqrt{q}} \sum_{l=0}^{q-1}e^{-i\phi_l /2} \mathcal{F}^{\pm}_l e^{-i\left(\hat{J}_z + \frac{\hat{\sigma}_z}{2} \right)\phi_l} \end{eqnarray} is a unitary operator that is a sum of unitary displacements $e^{-i\left(\hat{J}_z + \frac{\hat{\sigma}_z}{2} \right)\phi_l}$.

The explicit value of $\mathcal{F}_l^{\pm}$ is \begin{equation}  \mathcal{F}_l^{\pm} = e^{\mp \frac{i2\pi p}{q} \frac{N}{2} \left( \frac{N}{2} + 1 \right)} \frac{1}{\sqrt{q}} \sum_{m=0}^{q-1} e^{i \left[ \phi_l m \pm \frac{p}{q}\pi m (m-1) \right]} . \label{eq:gausssum} \end{equation} Sums of the kind in (\ref{eq:gausssum}) are known as \emph{generalized Gauss sums} \cite{Ber-81} and can be calculated for various values of $p$ and $q$. For simplicity we take $p=1$. In this case \cite{Ber-81} \begin{eqnarray}   \mathcal{F}_l^{\pm} &=& e^{\mp \frac{i2\pi}{q} \frac{N}{2} \left( \frac{N}{2} + 1 \right)}  e^{\pm i\frac{\pi}{4}} e^{\mp \frac{i\pi}{4} \frac{(2l\mp1)^2}{q} } \quad (q \: \mbox{odd}) , \\   \mathcal{F}_l^{\pm} &=& e^{\mp \frac{i2\pi}{q} \frac{N}{2} \left( \frac{N}{2} + 1 \right)} e^{\pm i\frac{\pi}{4}} e^{\mp \frac{i\pi}{4} \frac{(2l\mp1+1)^2}{q} } \quad (q \: \mbox{even}) . \end{eqnarray} 

\begin{figure}[]
\centering
\subfigure[]{
    \includegraphics[width=40mm]{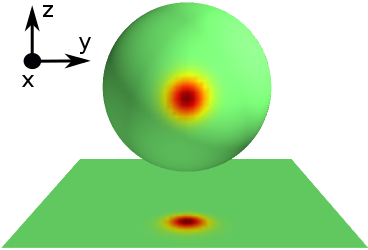}
}
\subfigure[]{
	\includegraphics[width=40mm]{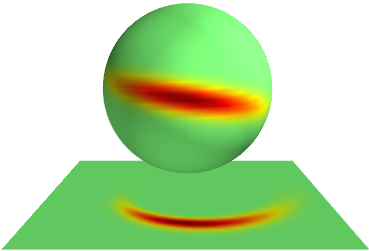}
}
\subfigure[]{
	\includegraphics[width=40mm]{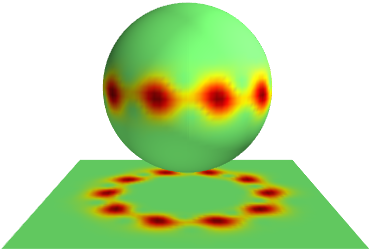}
}
\subfigure[]{
	\includegraphics[width=40mm]{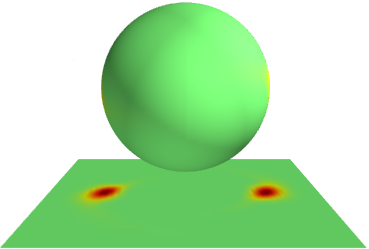}
}
\subfigure[]{
	\includegraphics[width=40mm]{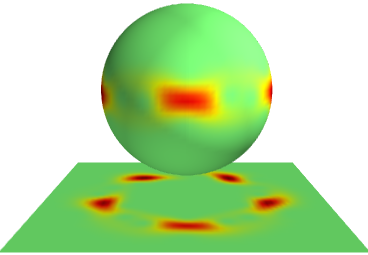}
}
\subfigure[]{
	\includegraphics[width=40mm]{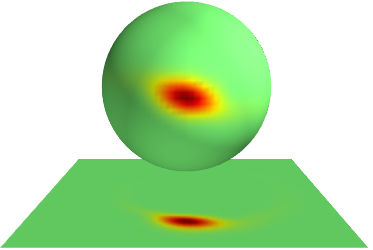}
}
\caption[]{(Color online). $Q$ functions [Eq. (\ref{eq:Q})] and their stereographic projections at various times for the (exact) state $\ket{\Psi^+ (t)}$ with $N=100$. (a) The initial state $\ket{\Psi^+ (0)} = \left[ \frac{1}{\sqrt{2}}\left( \ket{\uparrow} + \ket{\downarrow} \right) \right]^{\otimes N+1}$, a spin-coherent state. (b) The state $\ket{\Psi^+ (T/50)}$. (c) The state $\ket{\Psi^+ (T/10)}$. (d) The state $\ket{\Psi^+ (T/2)}$, a GHZ state. (e) The state $\ket{\Psi^+ (4T/5)}$. (f) The state $\ket{\Psi^+ (T)}$ at the revival time.} 
\label{fig:QN=100}
\end{figure}

The initial state $\ket{\Psi^{+}(0)}=\ket{\frac{\pi}{2}, \phi_0}_N \ket{\phi_0^+}$ is a spin-coherent state of the combined $(N+1)$-qubit system:

\begin{equation} \ket{\Psi^{+}(0)} = \left[ \frac{1}{\sqrt{2}}\left( \ket{\downarrow} + e^{-i\phi_0}\ket{\uparrow} \right) \right]^{\otimes (N+1)} .  \end{equation} In this case (ignoring global phase factors) the evolved state takes a particularly straightforward form:

\begin{eqnarray}   \ket{\tilde{\Psi}^{+} (T/q) } \hspace{-1.5mm}&\stackrel{(q \: \textup{odd})}{=}& \hspace{-0.5mm} \frac{1}{\sqrt{q}} \sum_{l=0}^{q-1}  e^{i\phi_l N/2} e^{- i\pi l(l - 1)/q} \ket{\frac{\pi}{2}, \phi_0 + \phi_l}_{N+1} , \\  \ket{\tilde{\Psi}^{+} (T/q) } \hspace{-1.5mm}&\stackrel{(q \: \textup{even})}{=}& \hspace{-0.5mm} \frac{1}{\sqrt{q}} \sum_{l=0}^{q-1}  e^{i\phi_l N/2} e^{- i\pi l^2/q} \ket{\frac{\pi}{2}, \phi_0 + \phi_l}_{N+1} . \end{eqnarray} This is a superposition of spin coherent states of the $N+1$ qubit system uniformly spaced around the equator of the Bloch sphere. This is in agreement with the results of the authors of \cite{Aga-97} and \cite{Chu-99} for the evolution of a spin-coherent state by a finite Kerr Hamiltonian. Taking $q=2$, for example, $\ket{\tilde{\Psi}^{+}(T/2)}$ is a Greenberger-Horne-Zeilinger (GHZ) state of the $N+1$ qubits. To visualize such states we plot in Fig. \ref{fig:QN=100} the Husimi $Q$ function \begin{equation} Q(\theta,\phi)=  \abssq{ \langle \Psi^{+}(t) | \theta,\phi \rangle_{N+1} } , \label{eq:Q} \end{equation} of the exact state $\ket{\Psi^{+} (t)}$ at various times for $N=100$ and for the initial state $\ket{\Psi^{+} (0)} = \left[ \frac{1}{\sqrt{2}}\left( \ket{\uparrow} + \ket{\downarrow} \right) \right]^{\otimes N+1}$. For very short times the initial spin-coherent state evolves to a spin-squeezed state \cite{Kit-93,Ma-11}, as shown in Fig. \ref{fig:QN=100}(b). At later times we see multiple-Schr\"{o}dinger-cat states [Figs. \ref{fig:QN=100}(c),(d),(e)]. We note that Schr\"{o}dinger-cat states can also be generated in a different parameter regime ($1/\sqrt{N} \ll |\zeta_0| \ll 1$) for the initial state, as discussed in \cite{Doo-13a}.  

\begin{figure}[ht]
\centering
    \includegraphics[width=90mm]{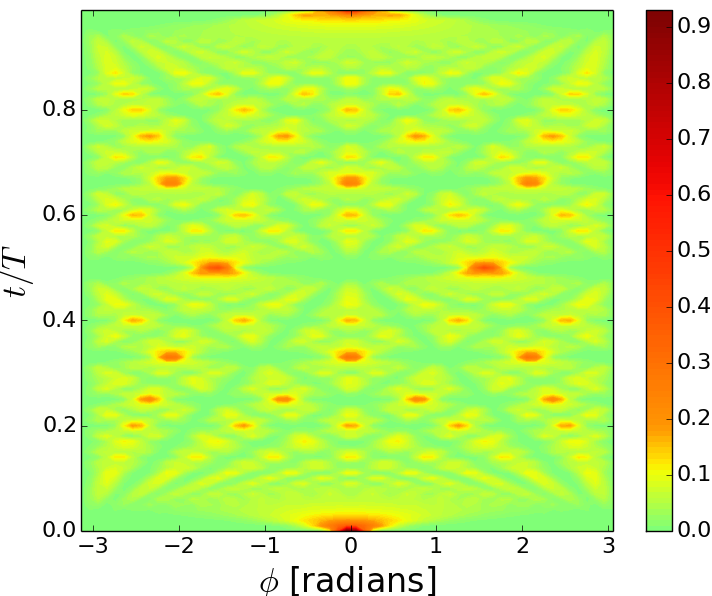}
    \caption{(Color online). Plotted is $Q\left( \frac{\pi}{2},\phi \right)$, the $Q$-function slice at $\theta = \pi/2$. Here $N=168$. We see a ``quantum carpet.''}
    \label{fig:carpet}
\end{figure}

Since the operator $J_z + \frac{\sigma}{2}$ commutes with our Hamiltonian (\ref{eq:H}), we know that it (and its powers) are conserved quantities of the system. This means that if the $Q$ function is initially a narrow distribution at the equator of the sphere, the $Q$ function is constrained to the equator at all times, as seen in Fig. \ref{fig:QN=100}. This suggests a concise visualization of the system dynamics by plotting $Q\left( \frac{\pi}{2},\phi \right)$ as a function of time and of $\phi$ (ignoring the variation in the polar angle $\theta$ that plays a less interesting role). The resulting pattern is shown in Fig. \ref{fig:carpet}. Such patterns are know as ``quantum carpets'' \cite{Kap-00,Ber-01}. At times $t=pT/q$ that are rational fractions of the period $T$ we see bright spots at the values of $\phi$ where there are spin coherent states.

\subsection{Small $N$}

Our approximations require that $|\zeta_0 |\approx 1$ (or $\theta_0\approx \pi/2$), $N\gg 1$ and $\lambda t \ll N$ (or, $t\ll T$). As illustrated in Fig. \ref{fig:precision}, the fidelity of the exact state to the approximate state gets worse as $N$ gets small or as $t$ gets close to $T$. Our $Q$-function plots show, however, that the qualitative features of the approximation are valid well outside of these parameter regimes.

In Fig. \ref{fig:carpet}, for example, it is clear that the multiple-Schr\"{o}dinger-cat states persist well beyond $t\ll T$. In Figs. \ref{fig:QN=100}(e) and \ref{fig:QN=100}(f) we plot the $Q$ functions for $N=100$ at $t = 4T/5$ and $t=T$, respectively. In Fig. \ref{fig:QN=100}(e) we see something qualitatively like a superposition of spin-coherent states, although the coherent states are distorted [the likely cause for the decrease in fidelity against the ideal superposition of spin-coherent states [Fig. \ref{fig:precision})]. 

\begin{figure}[]
\centering
\subfigure[]{
    \includegraphics[width=40mm]{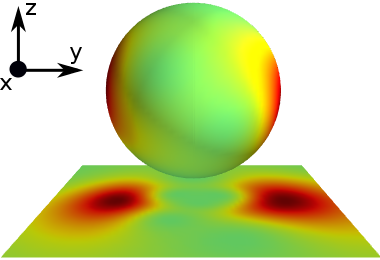}
}
\subfigure[]{
	\includegraphics[width=40mm]{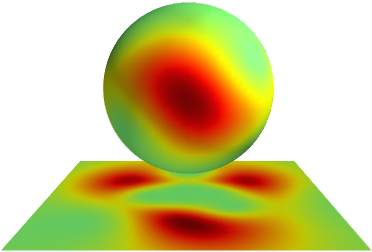}
}
\subfigure[]{
	\includegraphics[width=40mm]{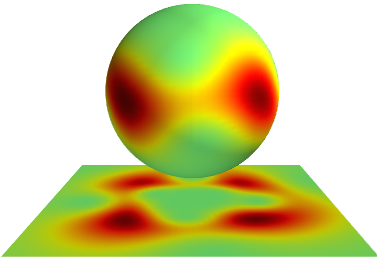}
}
\subfigure[]{
	\includegraphics[width=40mm]{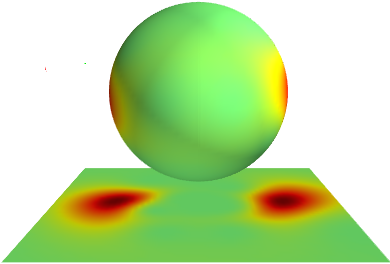}
}
\caption[]{(Color online). $Q$ functions (and their stereographic projections) at various times and for small values of $N$. Each time the initial state is $\ket{\Psi^+ (0)} = \left[ \frac{1}{\sqrt{2}}\left( \ket{\uparrow} + \ket{\downarrow} \right) \right]^{\otimes N+1}$. (a) $\ket{\Psi^+ (T/2)}$ for $N=8$. (b) $\ket{\Psi^+ (T/3)}$ for $N=10$. (c) $\ket{\Psi^+ (T/4)}$ for $N=16$;  (d) $\ket{\Psi^+ (T/2)}$ for $N=16$.} 
\label{fig:QNSmall}
\end{figure}

\begin{figure}[ht]
\centering
    \includegraphics[width=85mm]{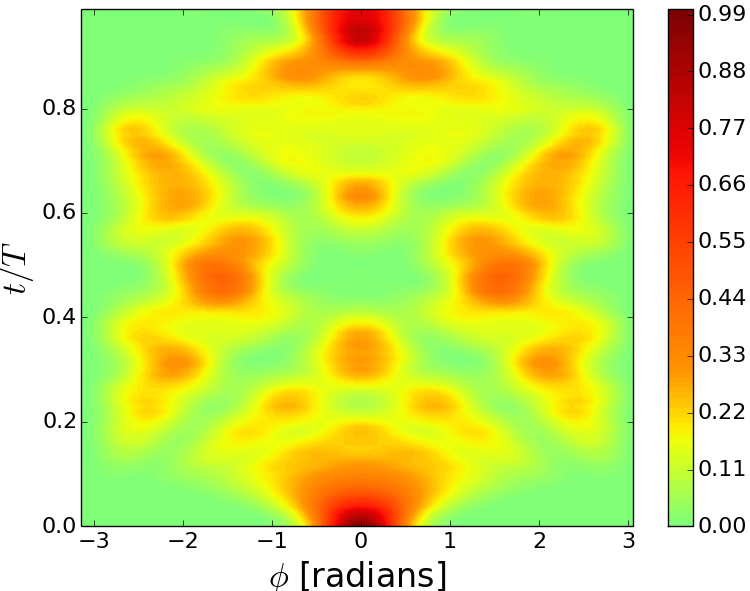}
    \caption{(Color online). Plotted is $Q\left( \frac{\pi}{2},\phi \right)$ for $N=16$. Even for this small value of $N$ the ``carpet'' pattern is conspicuous.}   
\label{fig:carpetsmallN}
\end{figure}

Similarly, although our approximation required that we assume $N\gg 1$, the $Q$ functions in Fig. \ref{fig:QNSmall} show that our approximation captures the qualitative features of the exact evolution of the system for moderately small values of $N$. It is clear from these plots that, although they are not superpositions of perfect spin-coherent states, they are superpositions of distorted spin-coherent states, still highly non-classical states. In Fig. \ref{fig:carpetsmallN} we plot the $\theta = \pi/2$ ``slice'' as a function of $\phi$ and of $t/T$ for $N=16$. The ``carpet'' pattern, although not as sharp as in Fig. \ref{fig:carpet}, is clearly recognizable. The states at $t=T/4$ and $t=T/2$ are plotted in Fig. \ref{fig:QNSmall}(c) and \ref{fig:QNSmall}(d), respectively. 

\section{Conclusion}

In \cite{Doo-13a} is was shown that a system of $N\gg 1$ qubits interacting with a single qubit via Hamiltonian (\ref{eq:H}) leads to collapse and revival analogous to that in the resonant Jaynes-Cummings model when the initial state of the $N$ qubits is a spin-coherent state $\ket{\zeta_0}_N$ with $ 1/\sqrt{N} \ll |\zeta_0| \ll 1$. Here we have shown that for an initial state in a different parameter regime, $| \zeta_0 | \approx 1$, the whole $(N+1)$-qubit system undergoes a different type of collapse and revival that includes fractional revivals: at rational fractions of the revival time the system is (approximately) in a multiple-Schr\"{o}dinger-cat state. The transition between the two different kinds of collapse and revival is made by changing the spin-coherent state parameter $|\zeta_0|$, or, equivalently, by changing $\theta_0$, the Bloch sphere polar angle for the initial spin-coherent state. 

The fractional revivals in the present case are similar to those found for the evolution of an $(N+1)$-qubit system by a finite Kerr Hamiltonian, the analogue of a Kerr Hamiltonian for an optical system \cite{Aga-97, Chu-99}. 

We emphasize that the initial spin-coherent state is, in principle, easily prepared since it consists of all qubits in the same pure state. By comparison, fractional revivals in the Jaynes-Cummings model requires an initial state with a sub-Poissonian photon number distribution \cite{Ave-92, Gor-93}. 




We also point out that the qualitative features of this approximation are valid even for moderately small values of $N$. An interesting application of this result might be the observation of fractional revivals in systems of few-qubits. Physical systems where this might be realized are interactions between superconducting qubits \cite{McD-05, Nis-07, Nee-10} for which the $\lambda \left( \sigma_+ \sigma_- + \sigma_- \sigma_+ \right)$ coupling between qubits is natural. Our interaction Hamiltonian (\ref{eq:H}) is composed of $N$ such equal interactions with the central qubit. Alternatively, a superconducting phase qudit \cite{Nee-09} (which emulates our $N+1$ dimensional $j=\frac{N}{2}$ subspace) might be coupled to a single superconducting qubit.


Another candidate system is any highly symmetric molecule that consist of $N$ spins equally coupled to a central spin. The trimethyl phosphite molecule, for example, has 9 $^{1}H$ spins, all equally coupled to a single $^{31}P$ spin \cite{Jon-09}. The tetramethylsilane molecule has 12 $^{1}H$ spins equally coupled to a single $^{21}S$ spin \cite{Sim-10}. These values of $N$ may be large enough to allow for the observation of fractional revivals.


An example of a system where Hamiltonian (\ref{eq:H}) may be realized for larger values of $N$ is the interaction of a superconducting flux qubit with a thin layer of NV centers \cite{Mar-10}. Rabi oscillations for such an interaction have been observed experimentally for $N\sim 10^7$ \cite{Zhu-11}. Squeezed states or Schr\"{o}dinger-cat states of the NV centers generated in this system may have applications in magnetic field sensing.






Future work will extend the model to $N$ qubits interacting with two qubits (an exactly solvable model), both from the perspective of the evolution of the $N$ qubits, and from the perspective of the decoherence and disentanglement of the two qubit system as it interacts with the $N$ qubits. Future work will also include a study of various forms of realistic decoherence to assess the possibility of experimentally observing fractional revivals, especially in systems of few qubits.

\bibliography{refs}

\appendix


\section*{Appendix A \label{app:a}}

Here we justify the approximations

 \begin{eqnarray} \left( \hat{J}_{-}\hat{J}_{+}  \right)^{-1/2} \hat{J}_{-} \ket{\zeta}_N &\approx & e^{-i\phi}\ket{\zeta}_N ,  \\ \left( \hat{J}_{+}\hat{J}_{-}  \right)^{-1/2} \hat{J}_{+} \ket{\zeta}_N &\approx & e^{i\phi}\ket{\zeta}_N , \end{eqnarray} when \begin{equation}  \frac{1}{\sqrt{N}} \ll \left| \zeta \right| \ll \sqrt{N} \quad \mbox{and} \quad  1 \ll N . \label{eq:appconds} \end{equation} Writing the spin-coherent state in its Dicke basis gives 
 
 \begin{equation} \ket{\zeta}_N =  \sum_{m=-N/2}^{N/2} C_m (\zeta) \ket{\frac{N}{2},m}_N ,\label{eq:SCS2} \end{equation} where \begin{equation} C_m (\zeta)  =  \binom{N}{\frac{N}{2}+m}^{1/2}\frac{\zeta^{\frac{N}{2}+m}}{\left( 1 + \abssq{\zeta} \right)^{N/2}} . \end{equation}

The operators $\left( \hat{J}_{-}\hat{J}_{+}  \right)^{-1/2} \hat{J}_{-}$ and $\left( \hat{J}_{+}\hat{J}_{-}  \right)^{-1/2}  \hat{J}_{+}$ act on the Dicke state $\ket{\frac{N}{2},m}$ as follows:

 \begin{eqnarray} \left( \hat{J}_{-}\hat{J}_{+}  \right)^{-1/2} \hat{J}_{-} \ket{\frac{N}{2},m} &= & \ket{\frac{N}{2},m-1} ,  \\   \left( \hat{J}_{+}\hat{J}_{-}  \right)^{-1/2} \hat{J}_{+} \ket{\frac{N}{2},m} &= & \ket{\frac{N}{2},m+1} , \end{eqnarray} so that we can write
 
  \begin{eqnarray} && \left( \hat{J}_{-}\hat{J}_{+}  \right)^{-1/2} \hat{J}_{-} \ket{\zeta_0}_N  \nonumber \\ && = \sum_{m=-N/2}^{\frac{N}{2}-1} \left[ \frac{\zeta\sqrt{\frac{N}{2}-m}}{\sqrt{\frac{N}{2}+m+1}}  \right] C_m (\zeta) \ket{\frac{N}{2},m}_N , \label{eq:app1} \\ && \left( \hat{J}_{+}\hat{J}_{-}  \right)^{-1/2} \hat{J}_{+} \ket{\zeta_0}_N \nonumber \\ && =  \sum_{m=-\frac{N}{2} + 1}^{\frac{N}{2}} \left[ \frac{\sqrt{\frac{N}{2}+m}}{\zeta\sqrt{\frac{N}{2}-m+1}}  \right] C_m (\zeta) \ket{\frac{N}{2},m}_N .  \label{eq:app2} \end{eqnarray} 
 
We expand the expressions in the square brackets in (\ref{eq:app1}) around the average values of $\frac{N}{2}\pm m$:

\small\begin{eqnarray}  && \quad \left[ \frac{\zeta\sqrt{\frac{N}{2}-m}}{\sqrt{\frac{N}{2}+m+1}}  \right]  \nonumber\\ && =   \left[ \frac{\zeta\sqrt{\frac{N}{2}-\bar{m}}}{\sqrt{\frac{N}{2}+\bar{m}}}  \right] \left( 1 - \frac{m-\bar{m}}{\frac{N}{2}-\bar{m}} \right)^{1/2}  \left( 1 + \frac{m-\bar{m}+1}{\frac{N}{2}+\bar{m}} \right)^{-1/2}  \nonumber \\ &&  \approx \left[ \frac{\zeta\sqrt{\frac{N}{2}-\bar{m}}}{\sqrt{\frac{N}{2}+\bar{m}}}  \right]\hspace{-1.5mm} \left( 1 - \frac{m-\bar{m}}{2\left( \frac{N}{2}-\bar{m} \right)} + \ldots \right) \hspace{-1.5mm} \left( 1 - \frac{m-\bar{m}+1}{2\left( \frac{N}{2}+\bar{m} \right)} + \ldots \right) . \nonumber \end{eqnarray} \normalsize The average value of $m$ with probability distribution $\abssq{C_m (\zeta)}$ is $\bar{m} = \expect{\hat{J}_z} = -\frac{N}{2}\left( \frac{1-\abssq{\zeta}}{1+\abssq{\zeta}} \right)$ so that \begin{equation} \frac{N}{2}-\bar{m} = \frac{N}{1+\abssq{\zeta}}\; ; \quad \frac{N}{2}+\bar{m} = \frac{N\abssq{\zeta}}{1+\abssq{\zeta}} \end{equation} and its standard deviation is $\Delta m = \Delta \hat{J}_z = \frac{\sqrt{N} |\zeta| }{1+\abssq{\zeta}}$. Since $m-\bar{m}$ will be of the order of $\Delta m$ we have

\small\begin{eqnarray}  && \quad \left[ \frac{\zeta\sqrt{\frac{N}{2}-m}}{\sqrt{\frac{N}{2}+m+1}}  \right]  \\ && \approx  \frac{\zeta}{|\zeta|} \hspace{-1.0mm} \left( 1 + \frac{|\zeta|}{2\sqrt{N}} + \ldots \right) \hspace{-1.5mm} \left( 1 - \frac{1}{2\sqrt{N}|\zeta| } - \frac{1}{2N\abssq{\zeta}} - \frac{1}{2N} + \ldots\right) . \nonumber \end{eqnarray}\normalsize When \begin{equation}  \frac{1}{\sqrt{N}} \ll \left| \zeta \right| \ll \sqrt{N} \quad \mbox{and} \quad  1 \ll N , \end{equation} we have

\begin{equation}  \left[ \frac{\zeta\sqrt{\frac{N}{2}-m}}{\sqrt{\frac{N}{2}+m+1}}  \right]  \approx \frac{\zeta}{|\zeta|} = e^{-i\phi} . \label{eq:appapprox1} \end{equation} Similarly, for the square bracket in $(\ref{eq:app2})$ we have

\begin{equation}  \left[ \frac{\sqrt{\frac{N}{2}+m}}{\zeta\sqrt{\frac{N}{2}-m+1}}  \right] \approx \frac{|\zeta|}{\zeta} = e^{i\phi} . \label{eq:approx2} \end{equation} Now, we would like to show that the contributions due to the terms $C_{N/2}$ and $C_{-N/2}$ that are missing in (\ref{eq:app1}) and (\ref{eq:app2}), respectively, are negligible. First, \begin{equation} \abssq{C_{N/2}} = \left( \frac{\abssq{\zeta}}{1+\abssq{\zeta}} \right)^N = \left( \frac{1}{1+\frac{1}{\abssq{\zeta}}} \right)^N . \end{equation} Since \begin{equation} \abssq{\zeta}\ll N \implies \frac{1}{1+\frac{1}{\abssq{\zeta}}} \ll \frac{1}{1+\frac{1}{N}} \end{equation} we can say that 

\begin{equation}  \abssq{C_{N/2}} \ll \left( \frac{1}{1+\frac{1}{N}} \right)^N \stackrel{(N\gg 1)}{\approx} \frac{1}{e} . \label{eq:approx1b} \end{equation} Similarly, using $\frac{1}{N} \ll \abssq{\zeta}$,

\begin{equation}   \abssq{C_{-N/2}} = \left( \frac{1}{1+\abssq{\zeta}} \right)^N \ll \left( \frac{1}{1+\frac{1}{N}} \right)^N \approx \frac{1}{e} , \label{eq:approx2b}\end{equation} so that both $C_{N/2}$ and $C_{-N/2}$ are negligible. Combining (\ref{eq:appapprox1}) and (\ref{eq:approx1b}) we have

\begin{equation}   \left( \hat{J}_{-}\hat{J}_{+}  \right)^{-1/2} \hat{J}_{-} \ket{\zeta}_N \approx e^{-i\phi}\ket{\zeta}_N . \end{equation} Combining (\ref{eq:approx2}) and (\ref{eq:approx2b}) we have

\begin{equation}  \left( \hat{J}_{+}\hat{J}_{-}  \right)^{-1/2} \hat{J}_{+} \ket{\zeta}_N \approx  e^{i\phi}\ket{\zeta}_N , \end{equation} as required.

\begin{widetext}

\section*{Appendix B \label{app:b}}

We truncate the Hamiltonian $\hat{H}^{\pm} = \pm \lambda \sqrt{\hat{J}^2 } \: \sum_{k=0}^{\infty} A_k \hat{M}^k$ to the first two terms so that $\hat{H}_\textup{trunc}^{\pm} = \pm \lambda \left( \sqrt{ \hat{J}^2  } - \frac{ \hat{J}_z^2 +  \hat{J}_z \otimes \hat{\sigma}_z }{ 2 \sqrt{ \hat{J}^2  }} \right)$. This is a good approximation if the fidelity $\left| \bra{\Psi^{\pm}(0)} e^{it\hat{H}_\textup{trunc}^{\pm}} e^{-it\hat{H}^{\pm}} \ket{\Psi^{\pm}(0) } \right|$ is close to unity.

\begin{eqnarray} \left| \bra{\Psi^{\pm}(0)} e^{it\hat{H}_\textup{trunc}^{\pm}} e^{-it\hat{H}^{\pm}} \ket{\Psi^{\pm}(0) } \right| =  \left| \bra{\Psi^{\pm}(0)} \exp\left[ \mp it\lambda \sqrt{\hat{J}^2} \sum_{k=2}^{\infty} A_k \hat{M}^k \right] \ket{\Psi^{\pm}(0) } \right| \\ = \left| \sum_{m=-N/2}^{N/2} \abssq{C_m} \bra{\phi_0^{\pm}} \exp\left[ \mp i t \lambda \sqrt{\frac{N}{2}\left(\frac{N}{2} + 1 \right)} \sum_{k=2}^{\infty} A_k \left( \frac{m^2 + m\hat{\sigma}_z}{\frac{N}{2}\left(\frac{N}{2} + 1 \right)} \right)^k \right] \ket{\phi_0^{\pm}}  \right| \label{eq:fid} \end{eqnarray}




The binomial distribution $\abssq{C_m}$ has average the value $\bar{m} = \expect{\hat{J}_z} = -\frac{N}{2}\left( \frac{1-\abssq{\zeta}}{1+\abssq{\zeta}} \right)$ and standard deviation $\Delta m = \Delta \hat{J}_z = \frac{\sqrt{N} |\zeta| }{1+\abssq{\zeta}}$. In (\ref{eq:fid}) we write $m = \bar{m} + (m-\bar{m})$ and replace all occurrences of $m-\bar{m}$ with the standard deviation $\Delta m$. This is reasonable because the coefficient $\abssq{C_m (\zeta)}$ with \begin{equation}  \frac{1}{\sqrt{N}} \ll \left| \zeta \right| \ll \sqrt{N} \qquad \mbox{and} \quad 1\ll N, \end{equation} is small unless $m$ is in the range $\bar{m}\pm\Delta m$. Only terms in this range will contribute significantly to the sum in (\ref{eq:fid}). Expanding the exponential and only keeping the $k=2$ term to lowest order in time gives

\begin{eqnarray}  \left| \bra{\Psi^{\pm}(0)} e^{it\hat{H}_\textup{trunc}^{\pm}} e^{-it\hat{H}^{\pm}} \ket{\Psi^{\pm}(0) } \right| \approx \left| 1\mp \sum_{m=-N/2}^{N/2} \abssq{C_m} \frac{i\lambda t A_2}{\left[\frac{N}{2}\left( \frac{N}{2}+1 \right) \right]^{3/2}} \bra{\phi_{0}^{\pm}}\left[ \left(\bar{m}+\Delta m\right)^2 + (\bar{m} + \Delta m)\sigma_z \right]^2 \ket{\phi_{0}^{\pm}} \right| \nonumber \\ \approx  \left| 1\mp \frac{i\lambda t A_2}{\left[\frac{N}{2}\left( \frac{N}{2}+1 \right) \right]^{3/2}} \bra{\phi_{0}^{\pm}}\left[ \frac{N^2 (1-\abssq{\zeta})^2}{4(1+\abssq{\zeta})^2}+\frac{N\abssq{\zeta}}{(1+\abssq{\zeta})^2} -\frac{N^{3/2}|\zeta| (1-\abssq{\zeta})}{(1+\abssq{\zeta})^2} -\frac{N(1-\abssq{\zeta})\sigma_z}{2(1+\abssq{\zeta})} + \frac{\sqrt{N}|\zeta| \sigma_z}{1+\abssq{\zeta}} \right]^2 \ket{\phi_{0}^{\pm}} \right| \end{eqnarray}

All terms are negligible if 

\begin{equation} \lambda t \ll \frac{2}{N} \left( \frac{1+\abssq{\zeta}}{1-\abssq{\zeta}} \right)^4 \quad \mbox{and} \quad \lambda t \ll \frac{N(1+|\zeta|^2)^4}{8|\zeta|^4}. \end{equation} For $|\zeta|=1$ this simplifies to the condition that $\lambda t \ll N$.

\end{widetext}



\end{document}